\documentclass[11pt, reqno]{article}
\usepackage{geometry}
\usepackage{amssymb}
\usepackage{setspace}
\usepackage{amsmath}
\usepackage{amsthm}
\usepackage{fancyref}
\usepackage{fancyhdr}
\usepackage{authblk}
\usepackage{cite}

\newcommand{\la}{\langle}
\newcommand{\ra}{\rangle}

\newcommand{\n}{\hat{n}}
\newcommand{\Hil}{\mathcal{H}}

\newtheorem{defn}{Definition}
\newtheorem{remark}{Remark}

\newtheorem{thm}{Theorem}
\newtheorem{cor}{Corollary}
\newtheorem{lem}{Lemma}
\newtheorem{prop}{Proposition}

\title{Anticoherent Subspaces}

\author[1]{Rajesh Pereira}

\author[1,2,3]{Connor Paul-Paddock}

\affil[1]{\emph{Department of Mathematics and Statistics, University of Guelph, Guelph, Ontario}}
\affil[2]{\emph{Institute for Quantum Computing, University of Waterloo, Waterloo, Ontario}}
\affil[3]{\emph{Department of Combinatorics and Optimization, University of Waterloo, Waterloo, Ontario}}

\date{}

\begin{document}

\maketitle

\begin{abstract}
We extend the notion of anticoherent spin states to anticoherent subspaces. An anticoherent subspace of order $t$ is a subspace whose unit vectors are all anticoherent states of order at least $t$. We use Klein's description of algebras of polynomials which are invariant under finite subgroups of $SU(2)$ on $\mathbb{C}^2$ to provide constructions of anticoherent subspaces. We discuss applications of this idea to the entanglement of $n$ qubit symmetric states. Furthermore, we show a connection between the existence of these subspaces and the properties of the higher-rank numerical range for a certain products of spin observables. We also note that these constructions give us subspaces of spin states all of whose unit vectors have Majorana representations which are spherical designs of order at least $t$.
\end{abstract}

\medskip


\section{Introduction}\label{sec:intro}

In this paper we examine subspaces of the spin $s$-states all of whose unit vectors are \emph{anticoherent} states, a concept first considered in \cite{zimba2006anticoherent}. Loosely speaking, anticoherent states have certain low degree spin moments whose values are independent of the axis of rotation. In \cite{zimba2006anticoherent} the author studied the properties of these anticoherent states using the \emph{Majorana representation}. The Majorana representation of a spin-$s$ state was formulated in \cite{majorana1932atomi}, as a tool to represent a spin-$s$ state as a set of points on the sphere $\mathcal{S}^2 \in \mathbb{R}^3$. It was noted in \cite{zimba2006anticoherent} that the Majorana representations of anticoherent states seemed to have a considerable amount of geometric structure. Another investigation \cite{crann2010spherical}, found a connection between the geometry of the Majorana representations for anticoherent states and so called \emph{spherical designs}, a concept introduced in \cite{delsarte1977spherical}. The authors even conjectured the equivalence of the Majorana representations for anticoherent states with these spherical designs \cite{crann2010spherical}. This conjecture, was however disproved in \cite{bannai2011note}, through constructions of counterexamples in both directions. The authors of \cite{crann2010spherical}, nonetheless additionally demonstrated that states which have a Majorana representation which is invariant under the orbit of a finite subgroup $G$ of $SO(3)$ are anticoherent states, whose order depends on $G$. It is this connection between the fixed point sets of finite symmetry subgroups and anticoherent states that have lead us to the results of this paper.\\

In the remainder of this introduction, we give a sketch of the main ideas and results in each section; full details can be found in the section.
We recall from \cite{zimba2006anticoherent}, that a spin-$s$ state is anticoherent of order $t$ if for a subset of spin observables $\mathcal{O}$ the $k$th power of the \emph{expectation} about the axis $\n$, $\la(\n \cdot \mathcal{O})^k\ra$ is independent of the unit vector $\n$ for all $k\leq t$ moments. An \emph{anticoherent subspace} is a subspace of the spin-$s$ Hilbert space for which every unit vector is an anticoherent state of order $t$. In this paper, we describe how one can construct these subspaces, and demonstrate that these subspaces arise in fairly low dimensions. \\

In Section \ref{sec:spin_states}, we review the Majorana representation for spin-$s$ states. In essence, the Majorana representation associates each spin-$s$ pure state with a set of $2s$ points on the sphere $\mathcal{S}^2$. The construction of the Majorana representation starts by first mapping the set of spin-$s$ states linearly to the set of complex polynomials of degree $2s$ or less. The representation is then attained by stereographically projecting the roots of the Majorana polynomial onto the sphere $\mathcal{S}^2$.\\

In Section \ref{sec:inva_polys}, we discuss the work of Felix Klein \cite{klein1956book}, on the generators of the algebra of homogenous polynomials in two variables which are invariant under some finite subgroup of $SU(2)$.  For each finite subgroup $G$ of $SU(2)$, Klein showed that the algebra of $G$-invariant polynomials is generated by three polynomials which he lists for each $G$. Using the Majorana representation and theorem 1 in \cite{crann2010spherical}, we can show that any spin state whose Majorana representation is a $G$-invariant polynomial is an anticoherent state.  Since the set of all $G$-invariant polynomials of degree less than or equal to $d$ form a vector space; the corresponding subspace of spin states consists entirely of scalar multiples of anticoherent states, and hence is an anticoherent subspace.  These constructions demonstrate that non-trivial examples of anticoherent subspaces exist for many values of spin.\\


In Section \ref{sec:symq_entanglement} we remark that a spin-$s$ state has the exact same mathematical structure as the symmetric product of $2s$ spin-$\tfrac{1}{2}$ states, allowing one to uniquely associate every Majorana polynomial with a symmetric product of qubits. When we stereographically project the roots of these polynomials to obtain the Majorana representation, we obtain an extension of the Bloch sphere representation. Hence the Majorana representation can uniquely describe pure symmetric multipartite qubit states. Many highly entangled states have a high degree of rotational symmetry in the states Majorana representation. This connection between anticoherence and entanglement was previously noted in \cite{aulbach2010geometric}. It follows that most anticoherent states and subspaces can be viewed as candidates for highly entangled multi-qubit systems.\\

In Section \ref{sec:hrnr}, we give a characterization of anticoherent subspaces in terms of the \emph{joint higher-rank numerical range} of the spin observables. The numerical range is a very useful tool in the study of observables and quantum state expectations and is a nice way to analyze these anticoherent subspaces at least for the cases where these joint higher rank numerical ranges are known.\\

We end in Section \ref{sec:sphere_des}, by considering subspaces all of whose unit vectors have Majorana representations that are spherical designs. Spherical designs of order $t$ are sets of $n$ points on the sphere $\{x_i\}$ which satisfy the equation

\begin{equation}
\sum_{i=1}^n f(x_i) = 0 \;,
\end{equation}

whenever $f$ is a \emph{harmonic polynomial} of degree $t$ or less. Even though the conjectured correspondence between anticoherent states and spherical designs has been proven to be false \cite{bannai2011note}, the results in \cite{crann2010spherical} show that the Majorana representations which arise from the $G$-invariant polynomials found by Klein are spherical designs.\\

\section{The Majorana Representation and Anticoherent States}\label{sec:spin_states}

The quantum spin-$s$ states span a $2s+1$ dimensional complex Hilbert space $\mathcal{H}_s$, with the natural orthonormal basis $\{|s,-s\ra, |s,-s+1\ra,\ldots,|s,s-1\ra, |s,s\ra\}$. We recall that $\mathbf{S}=\{S_x,S_y,S_z\}$ are the observables for the spin measurements along the $x$-, $y$-, and $z$-axis respectively. In the case of spin-$\frac{1}{2}$, these spin operators $\mathbf{S}=\frac{\hbar}{\sqrt{2}}\mathbf{\sigma}$, where $\mathbf{\sigma}=\{\sigma_x,\sigma_y,\sigma_z\}$ are the Pauli matrices. In the case of spin-$1$, the spin operators in the ordered basis
$\{ |1,-1\ra, |1,0\ra, |1,1\ra  \}$ are (omitting normalization)

\begin{equation}\label{eq:spin_obserables}
S_x =
\left(\begin{matrix}
0 & 1 & 0 \\
1 & 0 & 1 \\
0 & 1 & 0
\end{matrix}\right) \quad
S_y =
\left(\begin{matrix}
0 & -i & 0 \\
i & 0 & -i \\
0 & i & 0
\end{matrix}\right) \quad
S_z =
\left(\begin{matrix}
1 & 0 & 0 \\
0 & 0 & 0 \\
0 & 0 & -1
\end{matrix}\right)\;.
\end{equation}

We note that if $\n=(n_x, n_y, n_z)$ is any unit vector in $\mathbb{R}^3$, then the observable corresponding to spin about the $\n$-axis is $\n \cdot S=n_xS_x+n_yS_y+n_zS_z$.

\begin{defn}
A spin state is said to be anticoherent of order $t$ if  the expected value $\la(\n\cdot S)^k\ra$ is independent of the unit vector $\n$ for all natural numbers $k\leq t$.
\end{defn}


We now introduce the Majorana representation and its application to anticoherent states.  A vector spin-$s$ state can be written as state vector $|\psi\ra$ in the spin-$s$ Hilbert space $\Hil_s$

\begin{align}
|\psi\ra=\sum_{j=-s}^s a_j |s,j\ra \;.
\end{align}

To obtain the Majorana polynomial representation, $|\psi\ra$ is bijectively mapped to a complex polynomial of degree $2s$

\begin{align}
M_{|\psi\ra}(z)=\sum_{j=-s}^{s}(-1)^{j-s}a_{j}{2s \choose s+j}^{\frac{1}{2}}z^{s+j}\;.
\end{align}

We note that the variable $j$ runs from $-s$ to $s$ always in increments of length one; if $s$ is a half-integer, the sum is taken over all half-integers between $-s$ and $s$.  $M(z)$ has $2s$ roots in the complex plane, generating an unordered set of $2s$ points on the Riemann sphere $\mathcal{S}^2$ via stereographic projection from the north pole: $v:\mathbb{C}\cup\{\infty\}\rightarrow\mathbb{R}^3$,
\begin{equation*}v(z)=\frac{1}{|z|^2+1}(2\Re(z),2\Im(z),|z|^2-1),\hskip10pt v(\infty)=(0,0,1).\end{equation*}

This unordered set of $2s$ points is called the Majorana representation of $|\psi\ra$. When $s=\frac{1}{2}$, the Majorana representation reduces to the well known  Bloch sphere representation. The Majorana representation can best be pictured as a composition of the following correspondences.

\begin{equation}\label{eq:spinstate_isomorphism}
\text{spin-$s$ states} \leftrightarrow \text{roots of }M(z) \leftrightarrow \text{points in } \mathcal{S}^2\;.
\end{equation}

There is a close relation between many properties of the spin-$s$ state and the geometry of its Majorana representation. For instance, \emph{coherent} states are represented as a single point on the sphere with multiplicity $2s$. Eigenstates of $(\n\cdot S)$ with eigenvalue $j$ are represented by $s+j$ points at $\n$ and $s-j$ points at the antipode $-\n$.\\

While the Majorana polynomial is usually written as a complex polynomial is one variable, it often convenient to work with its \emph{homogenous} version.

\begin{align}
M^H_{|\psi\ra}(z,w)=\sum_{j=-s}^{s}(-1)^{j-s}a_{j}{2s \choose s+j}^{\frac{1}{2}} z^{s+j}w^{s-j}
\end{align}

Finally, we can define anticoherent subspaces as follows:

\begin{defn}
A subspace of the vector space of spin-$s$ states is called an anticoherent subspace of order $t$ if every unit element of the subspace is anticoherent of order $r\geq t$.
\end{defn}

The existence, construction and properties of these subspaces will be discussed in the subsequent sections of this paper.

\section{Invariant Polynomials of Finite Subgroups}\label{sec:inva_polys}

Zimba in \cite{zimba2006anticoherent}, proved that every spin state whose Majorana representation is a Platonic solid is anticoherent. This led to the natural question of whether there is a connection between the symmetry of the Majorana representation and the property of anticoherence.


\begin{thm}\cite{crann2010spherical} \label{cpk}
Let $G$ be a finite subgroup of $SO(3)$ then the orbit of $G$ on the two sphere is a Majorana representation of an anticoherent state whose order is bounded below by a positive integer constant depending only on $G$.  (This integer constant is one if $G$ is a dihedral group, two if $G$ is the tetrahedral group, three if $G$ is the octahedral group and five if $G$ is the icosahedral group).
\end{thm}

We note that if a spin state is invariant under a finite subgroup of $SU(2)$, its Majorana respentation is invariant under the corresponding finite subgroup of $SO(3)$ and its Majorana polynomial is invariant under the corresponding finite subgroup of $SU(2)$.  A characterization of the finite subgroups of $SU(2)$ and the structure of the set of polynomials which are invariant under these finite subgroups has been given by Klein \cite{klein1956book}. See the manuscript of Seade \cite{seade2005topology} for a modern account of these constructions.  Klein classified all finite subgroups of $SU(2)$. The finite subgroups consist of two infinite families of groups, the cyclic groups and the binary dihedral groups as well as three other groups: the binary tetrahedral group, the binary octahedral group and the binary icosahedral group.  Collectively, these groups are called the binary polyhedral groups.

\begin{thm}[Klein]
Let $G$ be a finite subgroup of $SU(2)$.  Then the set of homogenous polynomials invariant under $G$ is an algebra which is generated by three homogenous polynomials $h_1,h_2,h_3$.
\end{thm}

These homogenous polynomials are called the absolute invariants of $G$. Klein explicitly states the polynomials which generate the algebra of invariant polynomials for any given finite group of symmetries  in \cite{klein1956book} (See also \cite{dolgachev2009mckay,seade2005topology,toth2012finite} for other expositions of this material).

As examples, the absolute invariants of the binary dihedral group of order $4n$ are $h_1(z,w)=z^2w^2$, $h_2(z,w)=z^{2n}-2z^nw^n+w^{2n}$ and $h_3(z,w)=z^{2n+1}w-zw^{2n+1}$.  The absolute invariants of the binary tetrahedral group are $h_1(z,w)=z^5w-zw^5$, $h_2(z,w)=z^8+14z^4w^4+w^8$, $h_3(z,w)=z^{12}-33z^8w^4-33z^4w^8+w^{12}$. These and the absolute invariants of the other subgroups can be found in \cite{dolgachev2009mckay,toth2012finite}

The degrees of the invariant polynomials will be very important; we summarize this in the following table:

\begin{equation}
\begin{tabular}{|l|c|l|c|}
\hline
\textbf{Subgroup Name} & \textbf{$t$-anticoherence} & \textbf{$deg(h_1,h_2,h_3)$}\\ \hline
binary dihedral of order $4n$ & 1 & $(4,2n,2n+2)$  \\ \hline
binary tetrahedral & 2 & $(6,8,12)$  \\ \hline
binary octahedral & 3 & $(8,12,18)$ \\ \hline
binary icosahedral & 5 & $(12,20,30)$ \\ \hline
\end{tabular}
\end{equation}


\begin{remark}
Let $H_G$ be the set of polynomials which are invariant under $G$, a finite subgroup of $SU(2)$, then

\begin{align}
\widetilde{H}_G=span\bigg\{{h_1^u}{h_2^v}{h_3^w}\bigg\}_{u,v,w \in  \mathbb{N}\cup \{0\}}.
\end{align}

Furthermore $H_G$  is a graded algebra

\begin{equation}
H_G=\bigoplus_{d=0}^{\infty} H^d_G \;,
\end{equation}
where $H^d_G$ is the vector space of all invariant homogenous polynomials of degree $d$.
\end{remark}

\begin{remark}
Whenever we have nonnegative integer solutions for $u,v$ and $w$ of
\begin{equation}\label{majeq}
{u}\cdot deg(h_1)+{v}\cdot deg(h_2)+{w}\cdot deg(h_3)=2s,
\end{equation}
The polynomials $h_1^uh_2^vh_3^w$ corresponding to these solutions are a spanning set for a spin-$s$ anticoherent subspace.
\end{remark}

The number of distinct solutions which exist for a particular $s$ places an upper bound on the dimension of the spin-$s$ $t$-anticoherent subspace. The exact dimension can be calculated by enumerating the number of products of $h_1$, $h_2$ and $h_3$ that have the right degree and then checking for and eliminating linear dependencies all of which involve at least three separate terms.  This means that any two different polynomials of the form $h_1^uh_2^vh_3^w$ having the same degree will always be linearly independent but three such polynomials may not be. To find non-trivial anticoherent subspaces (i.e. those of dimension two or greater), we simply look for values of $s$ that give us at least two distinct nonnegative integer solutions for $(u,v,w)$ in equation \ref{majeq}. The problem of which numbers $s$ yield at least $k$ solutions for equation \ref{majeq} was considered in \cite{beck2002extension, beck2004formula}.

The binary dihedral group of order $8$ gives us a $1$-anticoherent subspace when spin $s$ is any integer other than one or three.  The binary tetrahedral group gives us $2$-anticoherent subspaces of dimension at least two when $s=6$ or $s$ is any even integer greater than or equal to $10$.  The binary octahedral group gives us $3$-anticoherent subspace of dimension at least two for $s=12,16,18,20,21,22$ and all integer values of $s$ greater than or equal to $24$.  The binary icosahedral group gives us a $5$-anticoherent subspace of dimension two for $s=30$ as well as for certain larger values of $s$.

To give a concrete example, let us find a basis for the $2$-anticoherent subspace of the spin-$6$ states obtained from the binary tetrahedral group. The degree twelve invariant polynomials are spanned by $h_1(z,w)^2=z^{10}w^2-2z^6w^6+z^2w^{10}$ and $h_3(z,w)=z^{12}-33z^8w^4-33z^4w^8+w^{12}$.  These are the Majorana polynomials of unique states, let us denote these states by $|\psi_1\ra$ and $|\psi_2\ra$.  The  subspace spanned by $|\psi_1\ra $ and $|\psi_2\ra $ is $2$-anticoherent.




\section{Symmetric Multipartite Qubit State Entanglement}\label{sec:symq_entanglement}

For collections of symmetric pure quantum states made up of $2s$ spin-$\frac{1}{2}$ states, the Majorana representation has two further properties which will play a role in our following discussion. Firstly we remark that applying a symmetric local unitary to the system of  $2s$ spin-$\frac{1}{2}$ states induces a rotation of its Majorana representation. Furthermore, when $s=\frac{1}{2}$ this representation is equivalent to the well-known Bloch sphere representation. Lastly, since any spin-$s$ state can be written as the symmetric product of $2s$ spin $\frac{1}{2}$ states, we can uniquely associate every Majorana polynomial with a symmetric product of qubit states $dim(\mathcal{H}_s)=2$.\\

 Recall a pure two level quantum state is a spin-$\tfrac{1}{2}$ state $|\phi\ra=\alpha|\frac{+1}{2}\ra+\beta|\frac{-1}{2}\ra$, were $|\alpha^2|+|\beta^2|=1$.
\begin{defn}
A multipartite symmetric $n$-qubit state is of the form
\begin{equation}\label{eq:symq_state}
|\phi^{(n)}\ra=\sum_{\sigma \in S_n}\left( \bigotimes\limits_{i=1}^n |\phi_{\sigma(i)}\ra\right)
\end{equation}
where each $\phi_i \in \mathcal{H}$ and $\sigma$ is a permutation of the symmetric group on $n$ objects $S_n$.
\end{defn}

Each state of the form \eqref{eq:symq_state} has a unique representation as a set of Majorana points $\left\{x_i\right\}_{i=1}^n$ on $S^2$ through the correspondence in \eqref{eq:spinstate_isomorphism}.\\

It has been shown that there is a relationship between the \emph{spread} of a multipartite qubit states Majorana representation and the entanglement of the state \cite{aulbach2010geometric,martin2010multiqubit}.  In fact by a result in \cite{leech1957equilibrium}, which we state below, one can argue that the states which have their Majorana representation maximally spread out tend to be invariant under certain finite subgroups of $SO(3)$, allowing us to obtain a class of states that are highly entangled.\\

We first remark that both entanglement and anticoherence appear to be properties which are maximized when the Majorana representation admits certain types of symmetry. The most obvious is that entanglement appears to maximized when the points in the Majorana representation are as spread out as possible from one another as noted in \cite{martin2010multiqubit}. However, the problem of maximizing distances between points on a sphere is far from trivial when one has no information about the underlying behaviour of the \emph{potential}.\\

The following result of Leech \cite{leech1957equilibrium} allows us to consider the problem of spread in terms of rotational symmetry. The relationship is due to the fact that sets of critical points for these potential functions are highly symmetrical.

\begin{thm}[Leech, \cite{leech1957equilibrium}]
The sets of critical points on $S^2$ which remain fixed under any potential are those which are invariant under non-trivial rotation about each point.
\end{thm}

Using this approach we consider a family of configurations that depend in no way on a potential. This allows us to think of them purely algebraically instead of with respect to a certain analytical characterization.

The set of rotations which leave these points invariant form a group under composition. We denote the set of configurations that are invariant under rotations about each point as the set of Leech configurations.  The Leech configurations are explicitly found in \cite{leech1957equilibrium}; there are finitely many and they largely consist of either the set of vertices of a Platonic solid or the set of midpoints of all edges of a Platonic solid.  We claim that the set of Leech configurations are candidates for high entanglement and these states are all anticoherent by theorem \ref{cpk}. From here the remarked connection between anticoherence and entanglement provides for the construction of candidate highly entangled multipartite qubit states.  Leech's result has been generalized considerably in \cite{cohn2007universally} where classes of  optimal configurations on higher dimensional spheres have been found.

\section{The Higher Rank Numerical Range and Anticoherent Subspaces}\label{sec:hrnr}

We begin by introducing the joint numerical range.

\begin{defn} Let $H=(H_1,H_2,...,H_m)$ be an $m$-tuple of Hermitian operators on a Hilbert space $\mathcal{H}$.  Then joint numerical range of $H$ is the subset of $\mathbb{R}^m$ defined as follows:\newline $\{ (\langle \phi |H_1|\phi \rangle,\langle \phi |H_2|\phi \rangle...\langle \phi |H_m|\phi \rangle): |\phi\rangle\in\mathcal{H}, \Vert |\phi \rangle \Vert=1 \}$.  \end{defn}

The joint numerical range is convex when $m\le 2$ or when both $m=3$ and $\dim(\mathcal{H})\geq 3$.  There are examples where the joint numerical range is not convex in other cases.

The joint rank-k numerical range numerical range was first defined in \cite{kribs2009research}.

\begin{defn}  let $H=(H_1,H_2,...,H_m)$ be an $m$-tuple of Hermitian operators on a Hilbert space $\mathcal{H}$.   Then joint numerical range of $H$ is the subset of $\mathbb{R}^m$ defined as follows:  $\Lambda_k(H)=\{((x_1,x_2,...,x_m): PH_jP=x_jP$ for some rank$-k$ projection $P$ and for all $i:1\le i\le m$. \end{defn}

The joint rank-1 numerical range is simply the joint numerical range.

\begin{defn} Let $V_{s,t}$ denote the vector space of trace zero $2s+1$ dimensional matrices spanned by all products of at most $t$ not necessarily distinct spin $s$ spin operators. \end{defn}

The vector space $V_{s,1}$ has a particularly simple structure.  The spin matrices $S=\{ S_x, S_y, S_z\}$ form a basis for $V_{s,1}$.

The following result follows almost immediately from the definition of an anticoherent state:

\begin{lem}  Let $|\phi \rangle$ be a spin $s$ pure state.  Then $|\phi \rangle$ is anticoherent of order $t$ if and only if $\langle \phi |A| \phi \rangle =0$ for all $A\in V_{s,t}$.    \end{lem}

We can rephrase this result in terms of the joint numerical range.

\begin{cor}  There is a spin $s$ pure state which is anticoherent of order $t$ if and only if the joint numerical range of any set of Hermitian matrices which forms a basis of $V_{s,t}$ contains the origin.\end{cor}

\begin{thm}  \label{hrmain} There is a $k$-dimensional $t$-anticoherent subspace of the spin $s$ states if and only if the joint rank $k$ numerical range of any set of Hermitian matrices which forms a basis of $V_{s,t}$ contains the origin.  \end{thm}

The joint higher rank numerical range of $S=\{ S_x, S_y, S_z\}$ is easy to calculate.

\begin{prop} If $s=\frac{1}{2}$, then $\Lambda_{1}(S)$ is the surface of the closed unit sphere.  If $s\geq 1$, then $\Lambda_{k}(S)$ is a closed solid sphere (with radius depending on $k$ and $s$) centered at the origin if $k\leq s+1$ and is empty otherwise. \end{prop}

 This allows us to use theorem \ref{hrmain} to find the maximal dimension of the $1$-anticoherent subspace for any spin.

\begin{cor} Let $s\geq 1$.  Then the largest $1$-anticoherent subspaces of the spin $s$ states have dimension $\lfloor s+1 \rfloor$.   \end{cor}

\section{Spherical Designs}\label{sec:sphere_des}

In what follows $\Omega_{n}=\{ x\in \mathbb{R}^{n}: \Vert x\Vert =1\}$ will denote the unit sphere in $\mathbb{R}^{n}$ endowed with the standard measure.

\begin{defn} \cite{delsarte1977spherical}  A spherical $t$-design on $\Omega_n$ is a finite set of points $x_1 ,x_2 ,...,x_k \in \Omega_n$ such that $\frac{1}{vol(\Omega_{n})} \int_{\Omega_n} p(x)dx = \frac{1}{k} \sum_{i=1}^{k} p(x_i)$ for all $n$ variable homogeneous polynomials $p(x)$ of degree less than or equal to $t$.  \end{defn}

Spherical designs were first introduced in \cite{delsarte1977spherical}.  Sobolev was the first to propose that orbits of finite subgroups of $O(n)$ be used as cubature for numerical integration over the surface of the $n$-sphere \cite{sobolev2006cubature}.

The connection between $t$-anticoherent states and spherical $t$-designs was noted in \cite{crann2010spherical}, where it was observed that spin-$1$ and spin-$\tfrac{3}{2}$ states where anticoherent of order $1$ if and only if their Majorana representations were spherical $1$-designs. These partial results lead the authors of \cite{crann2010spherical} to conjecture that a symmetric spin state is $t$-anticoherent if and only if its Majorana representation is a spherical $t$-design. However, it was shown in \cite{bannai2011note} that this conjecture was false in both directions.

The orbits of the finite non-Abelian subgroups of $SO(3)$ are all spherical designs.  An orbit of the tetrahedral group is a spherical design of order at least two, an orbit of the octahedral group is a spherical design of order at least three and an orbit of the icosahedral group is a spherical design of order five.

\begin{defn}
A subspace of the vector space of spin-$s$ states is called a spherical design subspace of order $t$ if the Majorana representation of every unit element of the subspace is a spherical design of order $r\geq t$.
\end{defn}

Our constructions of anticoherent subspaces all arise from finite symmetry groups and therefore all of the subspaces constructed in section three are also spherical design subspaces.

\section{Acknowledgements}

R.P. would like to thank NSERC for its support in the form of NSERC discovery grant 400550.  The authors are extremely grateful to the referee for catching an error in section three of the first draft of this paper.

\bibliographystyle{plain}

\bibliography{anticoherent}

\end{document}